\documentclass{svproc}
\usepackage{url}
\usepackage{graphicx}
\usepackage{comment}
\usepackage{multirow}
\usepackage{array}
\usepackage{amssymb}
\usepackage{tikz}
\usepackage{multirow}
\usepackage{enumitem,kantlipsum}
\usepackage{amsmath}
\usetikzlibrary{shapes.geometric, arrows.meta, positioning}
\newcolumntype{P}[1]{>{\centering\arraybackslash}p{#1}}
\newcolumntype{C}[1]{>{\centering\let\newline\\\arraybackslash\hspace{0pt}}m{#1}}
\usepackage{makecell}
\let\llncssubparagraph\subparagraph
\let\subparagraph\paragraph
\usepackage[compact]{titlesec}
\let\subparagraph\llncssubparagraph
\begin{document}
\mainmatter              % start of a contribution
\title{Enhancing Recommender Systems Using Textual Embeddings from Pre-trained Language Models}
\titlerunning{Enhancing RSs Using Textual Embeddings from Pre-trained LMs}  % abbreviated title (for running head)
\author{LE Ngoc Luyen, Marie-Hélène ABEL
}
\institute{Université de technologie de Compiègne, CNRS, Heudiasyc (Heuristics and\\ Diagnosis of Complex Systems), CS 60319 - 60203 Compiègne Cedex, France}
\authorrunning{NL LE et al.}

\maketitle              % typeset the title of the contribution

\begin{abstract}
Recent advancements in language models and pre-trained language models like BERT and RoBERTa have revolutionized natural language processing, enabling a deeper understanding of human-like language. In this paper, we explore enhancing recommender systems using textual embeddings from pre-trained language models to address the limitations of traditional recommender systems that rely solely on explicit features from users, items, and user-item interactions. By transforming structured data into natural language representations, we generate high-dimensional embeddings that capture deeper semantic relationships between users, items, and contexts. Our experiments demonstrate that this approach significantly improves recommendation accuracy and relevance, resulting in more personalized and context-aware recommendations. The findings underscore the potential of PLMs to enhance the effectiveness of recommender systems.
% We would like to encourage you to list your keywords within
% the abstract section using the \keywords{...} command.
\end{abstract}

\section{Introduction}

In recent years, the development of language models (LM) and pre-trained language models (PLM) has been foundational to the revolution in generative artificial intelligence (AI), as exemplified by systems like ChatGPT, Gemini, and others \cite{wang2022pre,achiam2023gpt,team2023gemini}. These advancements have significantly enhanced AI's capabilities in understanding and generating human-like text, and and they have profoundly influenced the field of natural language processing research \cite{ho2024survey}. In the context of recommender systems, PLMs offer a powerful tool for improving text-based tasks by providing deep semantic understanding and contextual insights \cite{zhao2023recommender}. %By integrating these models, recommender systems can better analyze and interpret textual data, such as user reviews, product descriptions, and comments, leading to more accurate and personalized recommendations.

Recommender systems are fundamental to many online platforms, enhancing user engagement and satisfaction by providing personalized suggestions from a wide range of products, services, or content \cite{le2023constraint,luyen_personalized}. However, traditional methods for building recommender systems typically rely on existing explicit information about users and items, and past user-item interactions or explicit user ratings, without leveraging the rich semantic information that can be captured through natural language representing of each user, or item information. By not incorporating natural language representations as input features, these methods miss out on the potential to enhance recommendation quality by integrating deeper contextual insights extracted from textual data associated with users or items.

To address these limitations, the integration of pre-trained language models presents a novel solution for enriching data inputs features by natural language representations. PLMs can be effectively utilized to extract rich semantic representations from textual data associated with items, user, and contexts. These models are highly effective at understanding natural language and can improve item and user profiles by using available textual data. Incorporating textual embeddings from PLMs into the recommendation process allows for a better understanding of the relationships between users, items and contexts, which in turn improves the quality of recommendations.

In this paper, we investigate enhancing RSs using textual embeddings from PLMs. By transforming user, item, and contextual data into natural language representations and generating high-dimensional embeddings using PLMs such as BERT \cite{devlin2018bert}, DistilBERT \cite{sanh2019distilbert}, and RoBERTa \cite{liu2019roberta}, we can enrich the input features for recommendation tasks. Our experimental results demonstrate that this approach significantly improves both the accuracy and relevance of recommendations compared to traditional methods.

The remainder of this paper is organized as follows: we describe the related work in Section \ref{section_relatedwork}. Following this, Section \ref{proposition} outlines our primary contributions, including the approach for data enrichment by using PLMs. In Section \ref{experiments}, we present experimental evaluations of our approach using real-word dataset. Finally, we conclude the paper in the last section.

\section{Related Work} \label{section_relatedwork}
We will first examine the related work on recommender systems, and then discuss the use of pre-trained language models to enhance these systems.
\subsection{Recommender System}
The Recommender System (RS) is traditionally characterized as an application designed to suggest the most relevant items to users by deducing or forecasting user preferences concerning an item. These systems leverage information about users, items, and their interactions to predict and recommend relevant content \cite{LU201512,le2021towards,le2023combining}.  There are six primary types of RS: Demographic-based, Content-based, Collaborative Filtering-basedn Knowledge-based , 	Context-aware, and Hybrid RSs  \cite{le2023improving}.
Each type of RS has data input and feature requirements. Demographic-based RSs use broad user demographic data, which is easy to implement but limits personalization. Content-based RSs rely on detailed item descriptions, providing high personalization but risking over-specialization without diverse item data. Collaborative Filtering RSs leverage user-item interactions, offering strong personalization but struggling with sparse data and the cold start problem. Knowledge-based RSs utilize structured item knowledge, needing extensive databases but not user interactions. Context-aware RSs include contextual data, enhancing relevance but adding complexity. Hybrid RSs merge data types to balance features and mitigate single-method limitations, increasing system complexity \cite{LU201512}. % The handling of input data for RS always play important role for developing models and give relevant recommendations.

The handling of input data for RSs always plays a crucial role in developing models and generating relevant recommendations. By effectively managing input data, it is possible to significantly enhance the accuracy, personalization, and user satisfaction of the recommendations provided. This underscores the importance of data quality and the strategies employed in data processing to ensure that RSs operate efficiently and effectively. With the advancement of LMs and PLMs, we can further leverage these developments to enhance data representation and enrich the information used in training models for RSs. These models enable deeper semantic analysis and a better understanding of text data, allowing for more relevant user profiles and item descriptions. In the next section, We examine related work on how PLMs can be applied to enhance RSs.

\begin{table}[h]
	\vspace{-0.5cm}
	\caption{The use of PLMs in Recommender Systems}
	\label{table_2_plm_recommenders}
	\centering
	\begin{tabular}{|P{1.6cm}|p{2.4cm}|p{2.8cm}|p{2.5cm}|P{1.0cm}|P{1.2cm}|}
		\hline
		\textbf{Works} & \textbf{Key Features}  & \textbf{Advantages} & \textbf{Drawbacks} & \textbf{PLM} & \textbf{RS} \\\hline
		BERT4Rec \cite{sun2019bert4rec} & Bidirectional self-attention for sequential recommendation & Highly effective at capturing sequential dependencies in user behavior. & High computational costs due to bidirectional processing. & BERT & Hybrid  \\\hline
		RecoBERT \cite{malkiel2020recobert} & Generates embeddings for both users and items & Enhances both user and item representation, improving personalization. & Requires large dataset to effectively train the embeddings. & BERT & Content-based  \\\hline
		U-BERT \cite{qiu2021u} & Pre-trains user representations for personalized recommendations & Improves recommendation accuracy by focusing on user-specific data pre-training. & Potential overfitting to the pre-training domain if not generalized well. & BERT & Hybrid \\\hline
		ONCE \cite{liu2024once} & Leverages both open-source and closed-source PLMs for recommendation & Enhances content understanding using a synergistic approach from multiple LM sources. & Complexity in integrating and managing different LM technologies. & BERT, LLaMA, GPT-3.5 & Content-based  \\\hline
	\end{tabular}
	\vspace{-0.8cm}
\end{table}

\subsection{Pre-trained Language Models}

A language model is defined as a probabilistic model of a natural language that predicts the next word or sequence of words in a given text based on the preceding context \cite{languagemodel}. Pre-trained language models are  language models that have been initially trained on large text corpora to learn general language patterns and contexts before being fine-tuned on specific tasks. These models, such as BERT \cite{devlin2018bert}, and RoBERTa \cite{liu2019roberta}, leverage deep learning architectures, primarily Transformers, to provide a robust understanding of language nuances, which can then be adapted to various natural language processing tasks without the need for training from scratch for each new application.

Recent years have seen a growing interest in leveraging PLMs to enhance RSs. The use of PLMs aims to better capture the semantic meaning of textual data associated with users and items, potentially leading to more accurate and relevant recommendations. A variety of approaches have been explored to integrate PLMs into RSs, as detailed in Table \ref{table_2_plm_recommenders}. These methods focus on embedding existing textual data from user reviews and item descriptions, which are utilized in content-based and hybrid models \cite{sun2019bert4rec,liu2024once}. Such advancements underscore the significant potential of PLMs to enhance the performance and effectiveness of RSs. Based on these advancements of PLMs, we will explore how to enhance the RS by leveraging PLMs in the next section.

\section{Our Approach to Data Enrichment} \label{proposition}
We first introduce the recommendation tasks of RSs, then we will present our approach to enhancing recommendations based on enriching textual data by using PLMs.

\subsection{Task Formulations}
In the context of RSs, the collected data typically consists of four fundamental components: (i) Item set $I = \{i_1, i_2, ..., i_n\}$, where $n$ is the total number of items;
	(ii) User set $U = \{u_1, u_2, ..., u_m\}$, where $m$ represents the total number of users; (iii) Context set $C = \{c_1, c_2, ..., c_k\}$, where $k$ is the number of contextual factors; (iv)
	Interaction set $R = \{r_{uic} | u \in U, i \in I, c \in C\}$, where $r_{uic}$ denotes the interaction between user $u$ and item $i$ under context $c$. In general, interactions can be represented in various forms, such as explicit ratings (1-5 stars), binary preferences (like/dislike), or implicit feedback (clicks, views, purchases). In a binary interaction scenario: $r_{uic} = 1$ indicates an observed interaction between user $u$ and item $i$ under context $c$. $r_{uic} = 0$ signifies the absence of an observed interaction.

The principal aim of a RS is to generate a ranked list of $N$ items tailored to the preferences of a specific user within a defined context. Our work concentrates on the task of data enrichment to improve the predictive accuracy of the RS. Accordingly, we begin with the structured tabular data concerning users $U$, items $I$, contexts $C$, and interactions $R$. The main task includes an initial transformation of this tabular data into natural language expressions that encapsulate comprehensive information about the users, items, and contexts. Following the data transformation phase, we proceed to an embedding generation phase, wherein PLMs are employed to create textual embeddings, or vector representations, of these natural language descriptions. This process of generating textual embeddings aims to enhance the representational richness of the data, thereby refining the system's ability to assess and predict item recommendations more accurately.

\subsection{Data Enrichment with Pre-trained Language Models}
\begin{figure}[ht]
	\vspace{-0.5cm}
	\centering
	\includegraphics[width=0.8\textwidth]{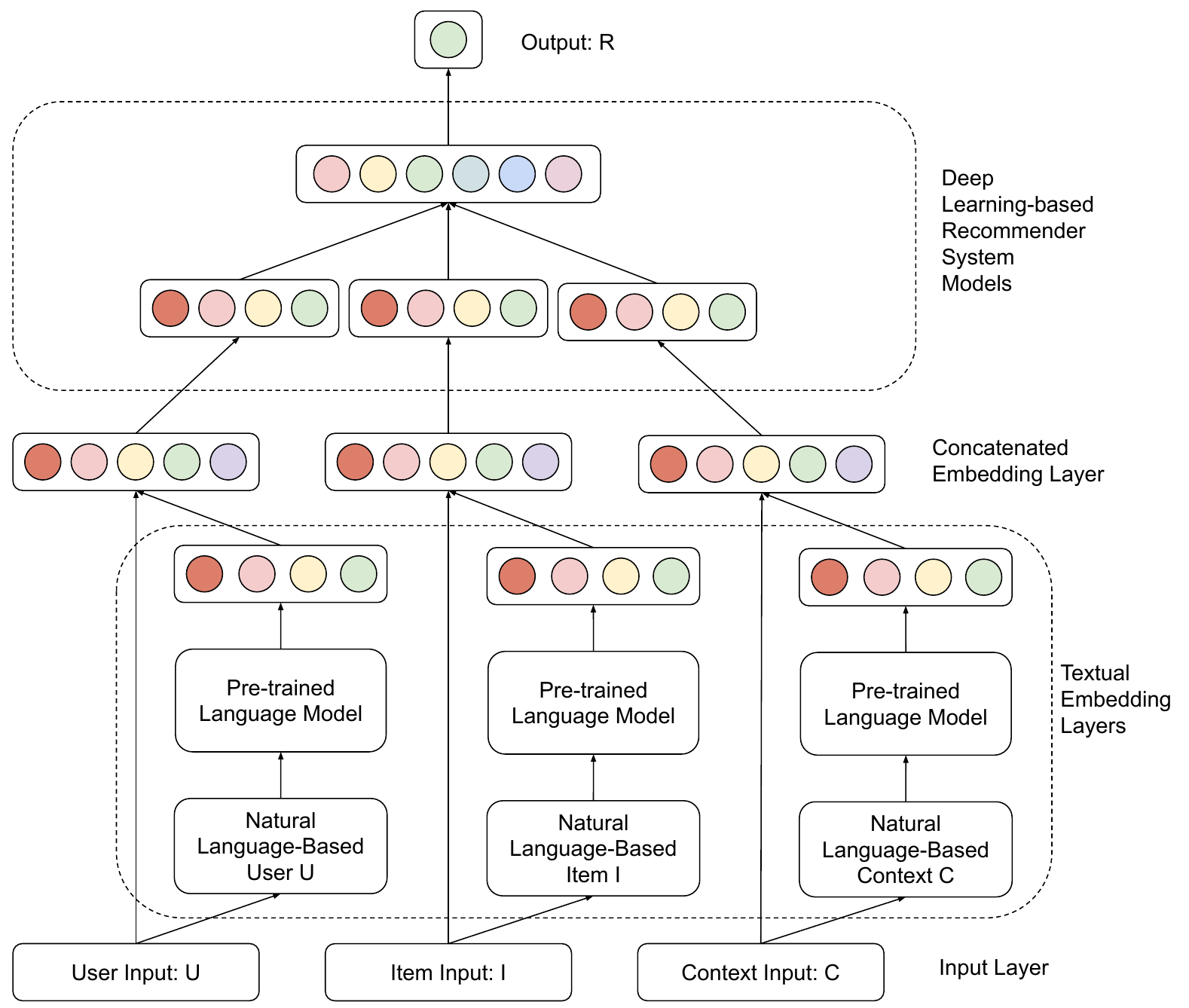}	
	\caption{The data enrichment architecture based on PLMs for RSs} \label{fig1}
\vspace{-0.5cm}
\end{figure}

Having established the task formulations, we now turn our attention to the architectural design that addresses these objectives. Our architecture leverages PLMs to enhance the recommendation process by incorporating enriched natural language data about users, items, and contextual information. This approach focuses on embedding text-based information to capture deeper semantic meanings often overlooked by traditional RSs. As illustrated in Figure \ref{fig1}, the architecture comprises five distinct component layers: Input Layer, Textual Embedding Layer, Concatenation Layer, Deep Learning-based Recommender System Models, and Output Layer.

\textbf{Input Layers:}
The input layer includes three types of data: User Input $U$, Item Input $I$, and Context Input $C$. User Input $U$ encompasses various data types including numerical (e.g., age, usage frequency), categorical (e.g., gender, occupation), and textual (e.g., profiles, reviews) information related to the user. Item Input $I$ similarly comprises numerical (e.g., price, ratings), categorical (e.g., genre, category), and textual (e.g., descriptions, titles) attributes and metadata associated with the item being recommended. Context Input $C$ incorporates numerical (e.g., timestamp, location coordinates), categorical (e.g., device type, season), and textual (e.g., current activity descriptions) contextual information that could influence the recommendation. This layer allows for the acquisition and integration of these diverse data types, potentially leading to more nuanced and context-aware recommendations.

\textbf{Textual Embedding Layers:}
This layer play an important role in our work by focusing on the enrichment of input data for Users, Items, and Contexts. Our approach involves a systematic transformation of structured data into natural language expressions, thereby enhancing the semantic richness of the input.
The process unfolds in two key steps: 
\begin{itemize}[wide, labelwidth=!, labelindent=0pt]
	\item \textit{Independent Processing}: Each input type (User, Item, and Context) undergoes independent processing to preserve its unique characteristics and relevance.
	\item \textit{Natural Language Transformation}: The structured data is then converted into coherent natural language expressions. This transformation serves to capture nuanced information that might be lost in traditional numerical or categorical representations.
\end{itemize}

For instance, a user profile typically represented in tabular format (e.g., {age: 20, gender: male, occupation: student, zip code: 60200}) is transformed into a natural language expression: ``\textit{The user is a 20-year-old male student residing in the Compi\`egne city with zip code 60200.}'' This approach offers several advantages: it preserves semantic relationships between attributes, incorporates domain knowledge into the language structure, and facilitates the capture of implicit information. By applying this process independently to each user, item, and context, we generate a corpus of natural language expressions that serve as enriched inputs for subsequent layers of the architecture."

The subsequent phase in our work leverages PLMs to generate vector representations for each natural language expression. These PLMs transform raw textual input into high-dimensional embeddings, effectively capturing the nuanced semantic relationships between the words. This process is underpinned by the PLMs' pre-training on massive corpora of natural language data, enabling them to encapsulate rich contextual and semantic information. As a result, we achieve User Embedding $U$, which captures the latent features from user-related text; Item Embedding $I$, where textual descriptions of items are transformed into dense vector representations; and Context Embedding $C$, wherein contextual features are encoded into embeddings.

\textbf{Concatenated Embedding Layer:}
In this layer, the embeddings from the user, item, and context -- derived both from tabular inputs and natural language expressions -- are concatenated into a unified representation. This layer consolidates all relevant information into a single, dense vector that effectively captures the interactions between the user, item, and context.

\textbf{Deep Learning-based Recommender System Models:}
The concatenated embedding is then inputted into a deep learning model specifically designed for recommendation tasks. This deep neural network processes the enriched embeddings to generate the final output, typically a predicted rating score $R$. The model is meticulously trained to optimize performance, utilizing default settings initially configured for generating recommendations of the deep learning model. 

Transformation of tabular data and leveraging PLMs to enrich data represent two key tasks addressed in this work. By integrating these tasks, our approach aims to significantly enhance the recommendation generation process, facilitating a more nuanced understanding of user preferences, item characteristics, and contextual factors. In the next section, we will conduct experiments on various deep learning models for the RS to demonstrate the performance of our data enrichment architecture.
\section{Experiments} \label{experiments}
To empirically validate the enhancements introduced by our data enrichment architecture, we begin by detailing the dataset used to evaluate the performance of our approach. We then describe the baseline deep learning models employed. Our goal is to measure the impact of our data enrichment architecture on the accuracy and efficiency of the RS.

\subsection{Dataset and Evaluation Measures} 
We conduct our experiments using the MovieLens ML-1M dataset\footnote{https://grouplens.org/datasets/movielens/}. This widely-used benchmark dataset is designed for the development and evaluation of RSs. It contains 1 million movie ratings from 6,042 users on 3,885 movies. The dataset includes both user demographic information: \{`\textit{age}', `\textit{gender}', `\textit{occupation}', `\textit{zip code}'\} and movie metadata \{`\textit{title}', `\textit{release year}', `\textit{genre}'\} making it ideal for testing models that rely on diverse input features. 

To evaluate performance, we use LogLoss and AUC (Area Under the ROC Curve) \cite{bradley1997use}. LogLoss measures the model's classification accuracy, with lower values indicating better performance. AUC assesses the model's ability to distinguish between positive and negative classes, with higher AUC values reflecting better discrimination between outcomes.
\subsection{Experiment Settings}
To evaluate the performance of our data enrichment architecture, we employed several well-established baseline models: 
\begin{itemize}[wide, labelwidth=!, labelindent=0pt]
	\item \textit{WideDeep}: This model captures feature interactions by including explicit interaction terms in a wide linear layer \cite{cheng2016wide}. 
	\item \textit{xDeepFM}: Combines a Compressed Interaction Network (CIN) with a deep neural network to capture both low- and high-order feature interactions \cite{lian2018xdeepfm}. 
	\item \textit{FiGNN} (Field-aware Graph Neural Networks): Leverages graph structures to model relationships between features and applies graph neural network techniques to learn interactions \cite{li2019fi}. 
	%\item \textit{AutoInt} (Automatic Feature Interaction Learning): Automates feature interaction learning using multi-view projections and attention mechanisms, reducing the need for manual feature engineering \cite{song2019autoint}. 
	\item \textit{DCNv2} (Deep Cross Network v2): Uses kernel products to capture complex interactions among multiple features simultaneously \cite{wang2021dcn}. 
	\item \textit{EulerNet}: Learns high-order feature interactions using spatial mappings based on Euler's formula to transform interactions into efficient representations \cite{tian2023eulernet}.
 \end{itemize}
The experiments were conducted using the PyTorch framework \cite{paszke2019pytorch} and the RecBole library \cite{zhao2022recbole}. To ensure optimal performance, we utilized the default hyperparameters specific to each model, as these have been pre-optimized. A batch size of 4096 was employed across all experiments. For the optimization of the training process, we implemented the Adam optimizer \cite{kingma2014adam}.
\begin{comment}
	
\begin{table}[h!]
	\centering
	\caption{Performance comparison of different models based on two metrics.}
	\begin{tabular}{|C{2.2cm}|C{1.1cm}|C{1.1cm}|C{1.1cm}|C{1.1cm}|C{1.1cm}|C{1.1cm}|C{1.1cm}|C{1.1cm}|}
		\hline
		\multirow{2}{*}{\textbf{Model}} & \multicolumn{2}{c|}{\textbf{raw}}& \multicolumn{2}{c|}{\textbf{\makecell{bert-based\\-uncased}}}   & \multicolumn{2}{c|}{\textbf{roberta-base}}  & \multicolumn{2}{c|}{\textbf{roberta-large}} \\\cline{2-9}
		& \textbf{AUC} & {\fontsize{6}{11}\textbf{LogLoss}}& \textbf{AUC} & {\fontsize{6}{11}\textbf{LogLoss}} & \textbf{AUC} & {\fontsize{6}{11}\textbf{LogLoss}}& \textbf{AUC} & {\fontsize{6}{11}\textbf{LogLoss}} \\\hline
		
		WideDeep & 0.8988 & 0.3189 & 0.8993 & 0.3144 & \textbf{0.8997} & \textbf{0.3141} & 0.8986 & 0.3156\\\hline
		xDeepFM & 0.8993 & 0.3167 & 0.8923 & 0.3287 & 0.8921 & 0.3244 & 0.8915 & 0.324 \\\hline
		FiGNN & 0.8961 & 0.3252 & 0.8937 & 0.3242 & \textbf{0.8974} & \textbf{0.3237}  & 0.8921 & 0.3293\\\hline
		AutoInt & \textbf{0.8914} & 0.3284 & 0.8876 & 0.3308 & 0.8907 & 0.3278 & 0.8905 & \textbf{0.3277} \\\hline
		DCNv2 & 0.8963 & 0.3273 & 0.8878 & 0.3348 & & &  0.8873  & 0.3280 \\\hline
		KD\_DAGFM & 0.8924 & 0.3262 & 0.8919 & 0.3265 & & 0.8901 & 0.3235\\\hline
		EulerNet & 0.8984 & \textbf{0.3227} & \textbf{0.8988} & 0.3272 & 0.8980 & 0.3228 & 0.8985 & 0.3320\\\hline
	\end{tabular}
	\label{tab:model_performance}
\end{table}
\end{comment}

\subsection{Experiments Results}
We employ various PLMs to generate textual embeddings for the natural language expressions of items, users, and contexts. These embeddings are then compared against the raw output of the baseline models. The comparison includes the following configurations: (i) \textit{raw}: the original results produced by the baseline models without textual embeddings from PLMs; (ii) \textit{bert-base-uncased}: results using BERT as the textual embedding model for natural language expressions \cite{devlin2018bert}; (iii) \textit{distillbert-base-uncased}: results using DistilBERT, a lighter and faster variant of BERT \cite{sanh2019distilbert}; (iv) \textit{roberta-base}: results using RoBERTa (base version) as the embedding model; and (v) \textit{roberta-large}: results using the larger RoBERTa model for potentially better capturing the nuances in the data \cite{liu2019roberta}.

\begin{table}[h!]
	\vspace{-0.5cm}
	\centering
	\caption{Performance comparison of different models.}
	\begin{tabular}{|C{2.5cm}|C{1.2cm}|P{1.5cm}|P{1.5cm}|C{1.5cm}|C{1.5cm}|C{1.5cm}|}
		\hline
		\multicolumn{2}{|c|}{\textbf{Metric}} & \textbf{\makecell{Wide\\Deep}} & \textbf{\makecell{xDeep\\FM}} & \textbf{\makecell{FiGNN}} & \textbf{\makecell{DCNv2}} & \textbf{\makecell{Euler\\Net}} \\ \hline
		
		\multirow{2}{*}{\textbf{raw}} & AUC & 0.8988 & 0.8948 & 0.8961  & 0.8896 & 0.8984 \\ \cline{2-7}
									& Logloss & 0.3189 & 0.3331 & 0.3252  & 0.3355 & 0.3227 \\ \hline
		
		\multirow{2}{*}{\textbf{\makecell{bert-based-\\uncased}}} & AUC & \textbf{0.8993} & 0.8945 & 0.8937  & 0.8913 & \textbf{0.8988} \\ \cline{2-7}
																& Logloss & \textbf{0.3144} & 0.3232 & 0.3242  & 0.3248 & 0.3272 \\ \hline
		
		\multirow{2}{*}{\textbf{\makecell{distilbert-\\based-uncased}}} & AUC & 0.8922 &0.8937 & 0.8960 &  0.8909 & 0.8987 \\ \cline{2-7}
																	& Logloss & 0.3236 & 0.3310 & \textbf{0.3213} &  0.3249 & \textbf{0.3202} \\ \hline
		
		\multirow{2}{*}{\textbf{\makecell{roberta-\\base}}} & AUC & 0.8921 & 0.8952 & \textbf{0.8974} &  0.8919 & 0.8985 \\ \cline{2-7}
														& Logloss & 0.3244 & 0.3231 & 0.3237 & 0.3256 &  0.3258 \\ \hline
		
		\multirow{2}{*}{\textbf{\makecell{roberta-\\large}}} & AUC & 0.8915 & \textbf{0.8959} & 0.8921 & \textbf{0.8956} &  0.8985 \\ \cline{2-7}
														& Logloss & 0.3240 & \textbf{0.3187} & 0.3293 & \textbf{0.3228} &  0.3320 \\ \hline
	
	\end{tabular}
	\label{tab:results}
	\vspace{-0.5cm}
\end{table}

The experimental results presented in Table \ref{tab:results} provide an overview of the performance of various RS models using different PLMs for data enrichment. In general, employing PLMs to enrich data through natural language representations leads to improved performance in most cases compared to raw data, with the most notable improvements observed in Logloss values. This demonstrates that PLMs contribute to reducing prediction errors, thereby enhancing the overall quality of predictions. The results also show varying performance patterns across models and PLMs. \textit{WideDeep} performs best with \textit{bert-based-uncased}, while \textit{xDeepFM} and \textit{DCNv2} benefit most from \textit{roberta-large}. \textit{FiGNN} achieves the highest AUC with \textit{roberta-base} and the lowest Logloss with \textit{distilbert-based-uncased}. \textit{EulerNet} demonstrates consistent performance across all PLMs, indicating its robustness to different input representations.

The effectiveness of PLMs varies across models: \textit{bert-based-uncased} consistently improves performance, especially for \textit{WideDeep}; \textit{distilbert-based-uncased} delivers mixed results but performs best on Logloss for \textit{FiGNN} and \textit{EulerNet}; \textit{roberta-base} enhances AUC, particularly for \textit{FiGNN}; and \textit{roberta-large} significantly boosts performance for \textit{xDeepFM} and \textit{DCNv2}. In some cases, there are trade-offs between AUC and Logloss improvements, as seen in \textit{EulerNet}'s performance with \textit{distilbert-based-uncased}. These results highlight the need to carefully match models with the right PLM to achieve the best outcomes, as different architectures benefit from different PLMs.

Overall, the experimental results show that using PLMs to enrich data generally improves RS performance. However, the effectiveness varies across different models and PLMs, highlighting the importance of choosing the right PLM - RS model combinations for optimal results. While the improvements are sometimes modest, they are consistent enough to demonstrate the potential of this approach in enhancing RSs.

\section{Conclusion}
In this paper, we explore enhancing recommender systems by using textual embeddings from PLMs. We develop a data enrichment architecture that leverages the transformation of traditional tabular data -- such as user attributes, item metadata, and contextual information -- into natural language representations. By employing advanced PLMs such as BERT, DistilBERT, and RoBERTa, we generate high-dimensional embeddings that capture deep semantic relationships and contextual nuances often overlooked by traditional RSs. Our experimental results show that these embeddings lead to notable improvements in both the accuracy and relevance of recommendations across various models. These findings suggest that careful selection of PLM - RS model combinations is critical for optimal performance. Looking ahead, future work could explore the scalability of this approach and its application to domain-specific contexts such as e-commerce, entertainment, and education, further validating its robustness and adaptability.

\begin{comment}

\begin{table}
\caption{This is the example table taken out of {\it The
\TeX{}book,} p.\,246}
\begin{center}
\begin{tabular}{r@{\quad}rl}
\hline
\multicolumn{1}{l}{\rule{0pt}{12pt}
                   Year}&\multicolumn{2}{l}{World population}\\[2pt]
\hline\rule{0pt}{12pt}
8000 B.C.  &     5,000,000& \\
  50 A.D.  &   200,000,000& \\
1650 A.D.  &   500,000,000& \\
1945 A.D.  & 2,300,000,000& \\
1980 A.D.  & 4,400,000,000& \\[2pt]
\hline
\end{tabular}
\end{center}
\end{table}
\end{comment}

%\bibliographystyle{splncs04}
\bibliographystyle{splncs03}
\bibliography{references}
\end{document}